\documentclass[prl,twocolumn,a4paper,superscriptaddress,showpacs]{revtex4}
\usepackage{graphicx,amsmath,bm,natbib, dsfont}

\def\text#1{\textrm{#1}}

\begin{document}

\title{Challenging preconceptions about Bell tests with photon pairs}
\author{V.~Caprara Vivoli}
\affiliation{Group of Applied Physics, University of Geneva, CH-1211 Geneva 4, Switzerland}
\author{P.~Sekatski}
\affiliation{Institut for Theoretische Physik, Universitat of Innsbruck, Technikerstr. 25, A-6020 Innsbruck, Austria}
\author{J.-D.~Bancal}
\affiliation{Centre for Quantum Technologies, National University of Singapore, 3 Science Drive 2, Singapore 117543}
\author{C.C.W.~Lim}
\affiliation{Group of Applied Physics, University of Geneva, CH-1211 Geneva 4, Switzerland}
\author{\\ B.G.~Christensen}
\affiliation{Department of Physics, University of Illinois at Urbana-Champaign, Urbana, Illinois 61801, USA}
\author{A.~Martin}
\affiliation{Group of Applied Physics, University of Geneva, CH-1211 Geneva 4, Switzerland}
\author{R.T.~Thew}
\affiliation{Group of Applied Physics, University of Geneva, CH-1211 Geneva 4, Switzerland}
\author{H.~Zbinden}
\affiliation{Group of Applied Physics, University of Geneva, CH-1211 Geneva 4, Switzerland}
\author{N.~Gisin}
\affiliation{Group of Applied Physics, University of Geneva, CH-1211 Geneva 4, Switzerland}
\author{N.~Sangouard}
\affiliation{Group of Applied Physics, University of Geneva, CH-1211 Geneva 4, Switzerland}
\affiliation{Department of Physics, University of Basel, CH-4056 Basel, Switzerland}

\begin{abstract}
Motivated by very recent experiments, we consider a scenario ``\`{a} la Bell'' in which two protagonists test the Clauser-Horne-Shimony-Holt (CHSH)  inequality using a photon-pair source based on spontaneous parametric down conversion and imperfect photon detectors. The conventional wisdom says that (i) if the detectors have unit efficiency, the CHSH violation can reach its maximum quantum value $2\sqrt{2}.$ To obtain the maximal possible violation, it suffices that the source emits (ii) maximally entangled photon pairs (iii) in two well defined single modes. Through a non-perturabive calculation of non-local correlations, we show that none of these statements are true. By providing the optimal pump parameters, measurement settings and state structure for any detection efficiency and dark count probability, our results give the recipe to close all the loopholes in a Bell test using photon pairs.
\end{abstract}
\date{\today}
\maketitle

\paragraph{Introduction.}
Many physicists are setting up challenging experiments to prove that non-locality is an element of the physical reality. The game is nonetheless simple. Two players, Alice and Bob, share a pair of entangled particles. Each chooses a measurement, $x$ for Alice and $y$ for Bob, among a set of two projectors represented by $\{x=0, x=1\}$ and similarly for $y.$ They get a binary result $\pm 1$, labelled $a$ and $b$ for Alice and Bob respectively. The game is repeated for as long as it is necessary to accurately estimate the probability distribution $p(ab|xy).$ Alice and Bob then compute the Clauser-Horne-Shimony-Holt (CHSH) \cite{Clauser69} value
\begin{equation}
\label{CHSH}
S=\sum_{x,y=0}^1 (-1)^{xy} \Big(p\left(a=b|xy\right)-p\left(a\neq b|xy\right)\Big) .
\end{equation}
If the CHSH inequality is violated i.e. if $S > 2$, Alice and Bob's correlations are non-local, namely their correlations cannot be reproduced by a strategy involving local hidden variables only. (Note that the CHSH inequality is the only relevant inequality in a scenario with two parties, two measurement settings and two results. In particular, the Clauser-Horne (CH) inequality \cite{Clauser74} is equivalent under the no-signaling assumption \cite{Brunner13}.) All the experiments realized so far point out that Nature is indeed non-local, but they all had loopholes.\\

There are two primary loopholes, the detection loophole and the locality loophole. The former uses undetected events to reproduce the observed correlations with a local model in which a conclusive result is given only when the measurement settings are in agreement with a predetermined strategy. It can thus be closed by guaranteeing that the number of undetected events is small enough \cite{Eberhard93}. The latter is closed when the measurement choice on Alice's side and the measurement result on Bob's side, and vice versa, are spacelike separated.  This guarantees that no local model in which the particles communicate the measurement settings they experience to choose the results accordingly, can explain the observed correlations.\\

\begin{figure}
\includegraphics[width=9cm]{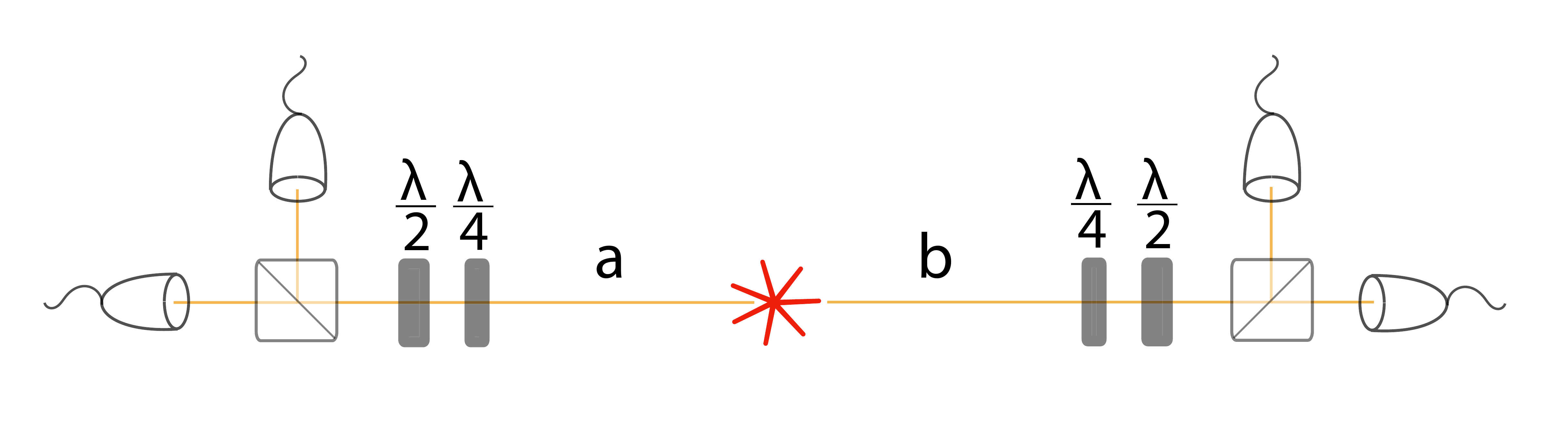}
\caption{A source (star) based on spontaneous parametric down conversion is excited e.g. by a pulsed pump and produced photon pairs entangled e.g. in polarization. The photons are emitted in correlated spatial modes $a$ ($b$). Each of them includes several temporal/frequency/spatial modes $a_k$ --- $b_k$, the number of temporal modes in the pulsed regime being given by the ratio between the pump duration and the photon coherence time for example. The photons emitted in $a$ ($b$) are sent to Alice's (Bob's) location where they are projected along an arbitrary direction of the Bloch sphere using a set of wave-plates, a polarization beamsplitter and two detectors. Each pump pulse triggers the choice of a measurement setting. The detectors are assumed to be non-photon number resolving with non-unit efficiency and dark counts.}
\label{fig1}
\end{figure}

The realization of a proper Bell test, i.e. without loopholes, would not only demonstrate that Nature is non-local, it would also open the way towards new applications. A detection-loophole free Bell test, for example, would allow one to realize device-independent randomness expansion where the size of an initial random bit string is made longer, the resulting randomness being guaranteed without the need to make assumptions about the internal working of the device used to extend the bit string (see \cite{Pironio10, Christensen13} for the first proof-of-principle experiments). In the same spirit, it would allow one to make device-independent quantum key distribution (see \cite{Acin07} for the principle and for example \cite{Gisin10, Sangouard11,Lim13} for experimental proposals). Independently of the purpose, the value of $S$ needs to be as close as possible to its maximum quantum value $2\sqrt{2}$. This makes the corresponding Bell test less demanding in terms of accumulated statistics to prove non-locality conclusively and more efficient regarding the randomness or the number of secret bits created per experimental run \cite{Brunner13}.\\

Although impressive results have been obtained with single atoms \cite{Rowe01, Matsukevich08, Hofmann12}, photons are natural candidates for loophole-free Bell tests. Actually, they have already been used to close both the locality loophole \cite{Aspect82, Tittel98, Weihs98} and the detection loophole \cite{Giustina13, Christensen13} even though these were in separate experiments. The basic setup exploits a photon pair source based on spontaneous parametric down conversion and photon detectors, as depicted in Fig. \ref{fig1}. The question that we address in this letter is what is the strategy that maximizes the CHSH violation in this specific scenario? Instead of using a perturbative approach, assuming e.g. that the source emits vacuum and from time to time one photon pair with a small probability, we present an exact calculation of correlations. This allows us to answer precisely and definitely the question above. We show for example, that the maximal value of $S$ is far from $2\sqrt{2}$ even if the detectors have a unit efficiency. This maximum is obtained through a multimode emission (Poissonian statistics) from non-maximally entangled states. Beyond the fact that these results go against collective intuition, they might significantly facilitate the realization of loophole-free Bell tests using photon pairs as they provide the method to follow to maximize the CHSH violation with non-photon number resolving detectors for any detection efficiency and dark count probability.\\

\paragraph{Modeling the pair source.} We first focus on the state produced by a photon pair source based on spontaneous parametric down conversion. Such a source produces photons in coupled modes, labelled by the bosonic operators $a_k$ and $b_k,$ the former is given to Alice, the latter to Bob. The subscript $k$ (which runs from $1$ to $N),$ means that Alice and Bob each receives several temporal/frequency/spatial modes. Furthermore, the photons are created in entangled states, e.g. in polarization, meaning that each mode splits into two orthogonal polarizations $a_k$ --- $a_{k,\bot}$ and $b_k$ --- $b_{k,\bot}.$ The Hamiltonian of the corresponding down-conversion process is $\mathcal{H}=i \sum_{k=1}^N(\chi a_k^\dag b_{k,\bot}^\dag - \bar\chi a_{k,\bot}^\dag b_k^\dag + h.c.)$ where $\chi$ and $\bar \chi$ are proportional to the non-linear susceptibility of the crystal and to the power of the pump \cite{footnotechi}. Their ratio determines whether maximally or non-maximally entangled states are produced. The exact expression of the state  produced by such a source $|\psi\rangle$ is obtained by applying the corresponding propagator $e^{-i \mathcal{H} t}$ on the vacuum $|\underbar{0}\rangle,$ as we are focusing on spontaneous emissions ($0$ is underlined to indicate that all modes are in the vacuum). As each mode $k$ is independent, i.e. two bosonic operators with different subscript $k$ commute, $e^{-i\mathcal{H}t}=\Pi_{k=1}^N e^{g a_k^\dag b_{k,\bot}^\dag- {\bar g} a_{k,\bot}^\dag b_k^\dag+h.c.}$ where $g=\chi t$ and $\bar g=\bar \chi t$ are the squeezing parameters for the coupled modes $a_k b_{k,\bot}$ and $a_{k,\bot} b_k$ respectively. Similarly, since $a_k b_{k,\bot}$ commute with $a_{k,\bot} b_{k},$ $e^{-i\mathcal{H}t}=\Pi_{k=1}^N U_k \bar{U}_k$ where $U_k = e^{g a_k^\dag b_{k,\bot}^\dag + h.c. },$ $\bar{U}_k=e^{-\bar g a_{k,\bot}^\dag b_{k}^\dag + h.c. }$ are squeezing operators. Finally, as the set $\{a_k^\dag b_{k,\bot}^\dag, a_k b_{k,\bot}, a_k^\dag a_k, b_{k,\bot}^\dag b_{k,\bot}\}$ is closed with respect to the commutator, $U_k = e^{T_g a_k^\dag b_{k,\bot}^\dag} C_g^{-(1+ a_k^\dag a_{k,\bot} +b_k^\dag b_{k,\bot})} e^{-T_g a_k b_{k,\bot}}$ \cite{Sekatski10}. $T_g$ $(C_g)$ stands for $\tanh(g)$ $(\cosh(g)).$ Using a similar formula for $\bar U_k,$ it is easy to show that
\begin{equation}
\nonumber
|\psi\rangle=(1-T_g^2)^{\frac{N}{2}} (1-T_{\bar g}^2)^{\frac{N}{2}} \Pi_{k=1}^N e^{T_g a_k^\dag b_{k,\bot}^\dag-T_{\bar g} a_{k,\bot}^\dag b_k^\dag} |\underbar{0}\rangle.
\end{equation}
Note that the number of modes $N$ is a tunable parameter. For $N=1,$ the photon statistic in each mode $a_k,  a_{k,\bot}, b_k, b_{k,\bot}$ corresponds to a thermal distribution whereas in the limit $N \rightarrow +\infty,$ it follows a Poissonian distribution. Moreover, the pair production in the modes $a_k^\dag b_{k,\bot}^\dag$ and $a_{k,\bot}^\dag b_k^\dag$ can be seen as two separate parametric processes that one pumps coherently e.g. by the same laser. The squeezing parameters $g$ and $\bar g$ and thus the amount of entanglement, can be tuned by controlling the pump power of each parametric process. \\

\paragraph{Modeling the photon detectors.} Let us now focus on the detectors. We consider photon detectors which do not resolve the photon number, do not distinguish the different modes $k$ and have non-unit efficiency $\eta.$ Formally, the event no-click corresponds to a positive operator $D_{\text{nc}}^a=\Pi_{k=1}^{N} C_L^{k\dag} T_{\text{nc}}^k C_L^k$ where $T_{\text{nc}}^k= |0\rangle_k\langle 0|$ is the projection operator on the vacuum for the mode $k$ and corresponds to an ideal non-photon-number-resolving detector and $C_L^k=e^{\gamma (a_k^\dag \ell_k-a_k \ell_k^\dag)} |0_{\ell_k} \rangle$ is the loss channel with $\eta=\cos^2\gamma$. $\ell_k$ is the initially empty mode whose coupling to $a_k$ is responsible for the loss. As $C_L^{k\dag} f(a_k) C_L^k=\langle 0_{\ell_k} | f(\sqrt{\eta} a_k + \sqrt{1-\eta} \ell_k)|0_{\ell_k}\rangle,$ and $|0\rangle_k\langle 0|= \-\ :e^{-a_k^\dag a_k}:$ where $: \-\ :$ is the normal order, we have $C_L^{k\dag} T_{\text{nc}}^k C_L^k=\-\ :e^{-\eta a_k^\dag a_k}:.$ Furthermore, since $:e^{(e^k-1) a_k^\dag a_k}:=e^{ka_k^\dag a_k}$ \cite{Collet88}, $D_{\text{nc}}^a$ can be written in a simple form as $D_{\text{nc}}^a=\Pi_{k=1}^{N} (1-\eta)^{a_k^{\dag}  a_k}.$ Note here that the detector dark counts (with the dark count probability $p_{\text{dc}}$) can be added by hand, as the probability for having no click requires all the $k$ modes to be empty and the absence of dark count i.e.
\begin{equation}
D_{\text{nc}}^a=(1-p_{\text{dc}})\Pi_{k=1}^{N} (1-\eta)^{a_k^{\dag}  a_k}.
\end{equation}
Analogously, the operator for a click $D_{\text{c}}^a$ is given by $\mathbf{1}-D_{\text{nc}}^a$ so that the detector is fully characterized by the positive operator valued measure $\{D_{\text{c}}^a, D_{\text{nc}}^a\}$.\\

\paragraph{Derivation of the probability distribution.} Now, we use the above models of the source and detectors to calculate the probability distribution $p(ab|xy)$ needed in the CHSH inequality. If the readers do not want to see the details on how they are derived, we invite them to go directly to the next section where the results are described. At each experimental run, Alice and Bob choose a measurement setting, i.e. they rotate the polarization of their modes
\begin{eqnarray}
\nonumber
a_k =&& \cos\alpha~A_k + e^{i \phi_\alpha}\sin\alpha~A_{k,\bot} \\
\label{rot_ak}
a_{k,\bot} = && e^{-i \phi_\alpha} \sin\alpha~A_k - \cos\alpha~A_{k,\bot}
\end{eqnarray}
(similarly for Bob with angles $\beta$ --- $\phi_\beta$ and the modes $B_k$ --- $B_{k,\bot}$) before they detect the modes $A_k, A_{k,\bot}, B_k, B_{k,\bot}$, see Fig. \ref{fig1}. They then look at their outcomes, i.e. they record which of their two detectors click. Locally, they can observe four different outcomes, either no click, one click in one of the two detectors or two clicks. Before we discuss the way to post-process the results, let us calculate for example the probability $p(\text{nc}_A)$ that Alice gets no click in $A$. It is obtained from
$
\text{tr } \left(D_{\text{nc}}^A |\psi_{\alpha,\beta}\rangle\langle \psi_{\alpha,\beta}|\right)
$
where tr stands for the trace over $A_k, A_{k,\bot}, B_k, B_{k,\bot}$ and $|\psi_{\alpha,\beta}\rangle$ is obtained by introducing the expressions of $A_k$ --- $A_{k,\bot}$ and $B_k$ --- $B_{k,\bot}$ given in (\ref{rot_ak}) in $|\psi\rangle.$ Note that $p(\text{nc}_A)=(1-p_{\text{dc}}) \left(\text{tr}~R^{A_k^{\dag}  A_k} |\psi^k_{\alpha,\beta}\rangle\langle \psi^k_{\alpha,\beta}| \right)^N$ with $|\psi_{\alpha, \beta}\rangle = \Pi_{k=1}^N |\psi^k_{\alpha, \beta}\rangle,$ $R=(1-\eta)$ and since the trace is cyclic $p(\text{nc}_A)=\left(1-p_{\text{dc}}\right) \left(\text{tr}~R^{\frac{A_k^{\dag}  A_k}{2}} |\psi^k_{\alpha, \beta}\rangle\langle \psi^k_{\alpha, \beta}| R^{\frac{A_k^{\dag}  A_k}{2}} \right)^N.$ Furthermore, from $x^{a^\dag a} f(a^\dag )= f(x a^\dag ) x^{a^\dag a}$ \cite{Sekatski10}, we have $R^{\frac{A_k^{\dag}  A_k}{2}} |\psi^k_{\alpha, \beta}\rangle=\left(1-T_g^2\right)^{\frac{1}{2}} \left(1-T_{\bar g}^2\right)^{\frac{1}{2}} e^{(A_k^\dag, A_{k,\bot}^\dag) M\left(\begin{matrix}B_k^\dag \\ B_{k,\bot}^\dag\end{matrix}\right)} |\underbar{0}\rangle$
with $M=\left(\begin{matrix}
R^{\frac{1}{2}}(T_gC_\alpha S_\beta^\star - T_{\bar g}S_\alpha^\star C_\beta) &
R^{\frac{1}{2}}(-T_gC_\alpha C_\beta - T_{\bar g}S_\alpha^\star S_\beta) \\
T_gS_\alpha S_\beta^\star + T_{\bar g}C_\alpha C_\beta &
-T_gS_\alpha C_\beta + T_{\bar g}C_\alpha S_\beta
\end{matrix}\right),$
$C_\alpha$ and $S_\alpha$ ($C_\beta$ and $S_\beta$) meaning $\cos \alpha$ and $e^{i\phi_\alpha}\sin \alpha$ ($\cos \beta$ and $e^{i\phi_\beta} \sin \beta$) respectively.
From the singular value decomposition of $M=U\left(\begin{matrix}
\lambda_1 &
0 \\
0 &
\lambda_2
\end{matrix}\right) V^\star,$ $R^{\frac{A_k^{\dag}  A_k}{2}} |\psi_k(\alpha, \beta)\rangle$ reduces to $\left(1-T_g^2\right)^{\frac{1}{2}} \left(1-T_{\bar g}^2\right)^{\frac{1}{2}} e^{\lambda_1 U A_k^\dag V B_k^\dag+ \lambda_2 U A_{k,\bot}^\dag V B_{k,\bot}^\dag} |\underbar{0}\rangle$ and we end up with the simple formula
\begin{eqnarray}
\label{proba_noclic_A}
&&p(\text{nc}_A)=(1-p_{\text{dc}})\left(\frac{\left(1-T_g^2\right) \left(1-T_{\bar g}^2\right)}{\left(1-\lambda_1^2\right)\left(1-\lambda_2^2\right)} \right)^N\\
\nonumber
&&=(1-p_{\text{dc}})\left(\frac{2}{2-\eta +\eta(C_{2g} C_\alpha^2 +C_{2\bar g} |S_\alpha|^2)} \right)^N.
\end{eqnarray}
Following the same line of thought, we can derive all the no-detection probabilities $p(\text{nc}_A~\&~\text{nc}_{A_\bot})$ etc... \\

To compute the CHSH value, Alice \& Bob need to bin their results, i.e. they have to choose a local strategy to assign the values $\pm 1$ to their four possible events. Among the 256 possible strategies to deal with the non-conclusive events, a simple strategy consists in assigning the value $-1$ to one of the results corresponding to  one click locally (the detectors $A$ ($B$) clicks whereas $A_{\bot}$ ($B_{\bot}$) does not) and $+1$ to the other events. Hence, $p(\text{-}1\text{-}1|xy)=\text{tr} \left((\mathbf{1}-D_{\text{nc}}^A)D_{\text{nc}}^{A_\bot}(\mathbf{1}-D_{\text{nc}}^B)D_{\text{nc}}^{B_\bot} |\psi_{\alpha,\beta}\rangle\langle \psi_{\alpha,\beta}|\right)$ and can be related to the no-detection probabilities derived previously through $p(\text{nc}_{A_{\bot}}~\&~\text{nc}_{B_{\bot}})-p(\text{nc}_A~\&~ \text{nc}_{A_{\bot}}~\&~\text{nc}_{B_{\bot}})-p(\text{nc}_{A_{\bot}}~\&~ \text{nc}_B~\&~\text{nc}_{B_{\bot}})+p(\text{nc}_{A}~\&~\text{nc}_{A_{\bot}}~\&~ \text{nc}_B~\&~\text{nc}_{B_{\bot}}).$ (See Appendix for the complete expressions.) Processing $p(\text{+}1\text{-}1|xy),$ $p(\text{-}1\text{+}1|xy)$ and $p(\text{+}1\text{+}1|xy)$ in a similar way makes it possible to optimize the CHSH value over the squeezing parameters ($g$ --- $\bar g$), the number of modes $N$ and the measurement settings for any detection efficiency and dark count probability. \\

\begin{figure}[ht!]
\includegraphics[width=8.5cm]{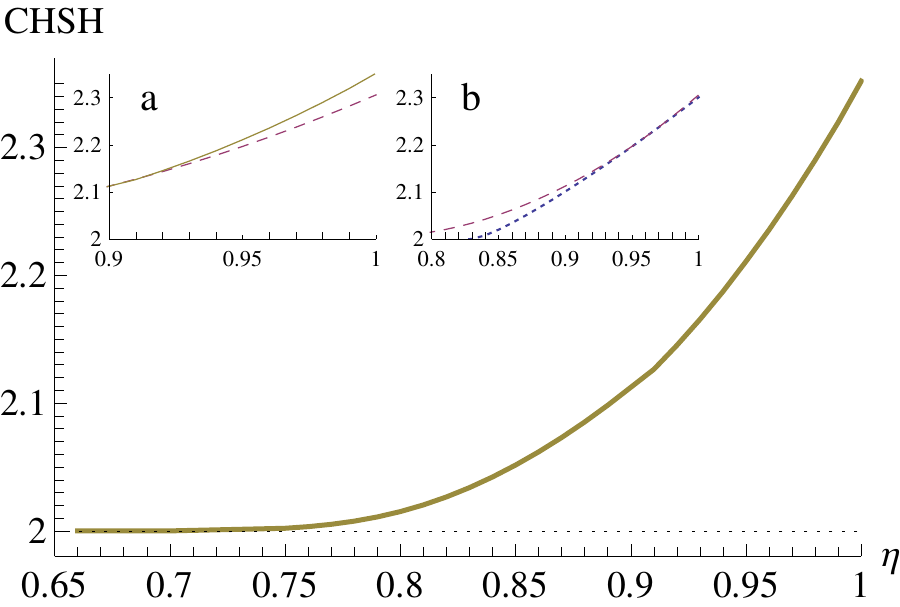}
\caption{CHSH values as a function of the detection efficiency $\eta.$ The full curve is optimized over the structure of the states produced by a spontaneous parametric down conversion source (the squeezing parameters and the number of modes), the measurement settings and the local strategy with which the outcomes are assigned to results $\pm 1.$ The dark count probability is set to zero. Inset a : the optimal violation (full curve) is compared to the CHSH value obtained by restricting the emission to a single mode i.e. thermal photon number statistics (dashed line). Inset b : Comparison between the CHSH values restricted to a single mode (dashed line) and the one obtained by focusing on the single mode case and restricting to maximally entangled states ($g=\bar g$, see dotted line). Note that the CH value can be deduced from the here shown CHSH value through $\frac{S-2}{4}.$  }
\label{fig2}
\end{figure}

\paragraph{Results.} The results are shown in Fig. \ref{fig2}. They have been obtained under the assumption that there is no dark count. Furthermore, we have checked that the strategy described before binning the four possible results locally is optimal. Therefore, the full curve of Fig. \ref{fig2} gives the maximal violation of the CHSH inequality that can be obtained in the scheme represented in Fig. \ref{fig1}. Several results deserve to be elaborated.\\

(i) The maximum CHSH value obtained with unit efficiency detectors is $\sim 2.35.$ This is very far from the maximal quantum value $2\sqrt{2}$ that can be obtained with any two-qubit states that are maximally entangled. The reason is that the photon pair source under consideration inevitably produces vacuum and multiple pairs. The vacuum leads to no-detections and the corresponding CHSH value is 2. Similarly, when many pairs are produced, the four detectors click. This also results in a CHSH value of 2. None of these events prevent the violation of the CHSH inequality but they reduce the observed violation. \\

(ii) The maximal violation is obtained when the number of modes tends to infinity (Poissonian statistics). For comparison, we can restrict the emission to be mono-mode (N=1) and still optimize the CHSH value over the squeezing parameters and the measurement settings. The corresponding results are given by the dashed line in the inset a of Fig. \ref{fig2}. The maximal CHSH-value is unchanged for efficiencies smaller than $\sim 91$\% but for higher efficiencies, the many-mode configuration favors larger violations. The intuition in the ideal case $\eta=1$ is that the relative probability for having a single pair is greater for the Poissonian distribution compared to the thermal distribution for a mean photon number around $1.$ However, for inefficient detectors, it becomes more difficult to have an intuition as in addition to the photon statistics, we have to take into account the detrimental effect of multi-photon events underlying losses.\\

(iii) In the case $\eta=1,$ the maximum violation is reached for a ratio $g/\bar g \sim 0.92,$ i.e. for non-maximally entangled states. For comparison, the dotted thin curve of the inset b in Fig. \ref{fig2} gives the optimal value when forcing the squeezing parameters to be the same (still in the monomode case). This shows that even in the monomode case, it is never optimal to use maximally entangled states, even when dealing with unit efficiency detectors. The intuition is that the two non-conclusive events (no-click and two clicks locally) both have an effect analogous to loss and we know from Eberhard \cite{Eberhard93} that non-maximally entangled states have a greater resistance to loss.\\

(iv) The minimum efficiency required to observe non-local correlations is 2/3. This is surprising at least at first sight, as this corresponds exactly to the minimum efficiency that is required to violate the CHSH inequality when dealing with two-qubit states. The intuition is that neither the vacuum nor the multiphoton events prevent the violation as each leads to $S=2.$ Hence, the CHSH inequality is violated as long as the two-qubit component (one photon pair exactly) leads to $S>2.$ To further decrease the required efficiency, we can investigate other Bell inequalities with more outputs and/or inputs. This provides work for future.\\

\paragraph{Conclusion \& Perspectives.}
We have presented an exact derivation of correlations in a scenario with two parties testing the CHSH inequality with a source based on spontaneous parametric down conversion. In particular, we have shown that the maximal CHSH value is $\sim2.35.$ This prevents the scenario drawn in Fig. \ref{fig1} to be used for device-independent quantum key distribution based on the CHSH inequality, as the secret key fraction goes to zero for $S \sim 2.43$ under the assumption of collective attacks \cite{Acin07}. 
Note that a higher violation can be obtained through amplification \cite{Gisin10} or non-linear filtering \cite{Sangouard11} at the price of increasing complexity. An alternative solution is based on conditioning, i.e. by heralding the creation of a single photon that is subsequently sent through a beam-splitter. By giving each output mode to Alice $\&$ Bob respectively, it has been shown that S can reach 2.68 \cite{BohrBrask13} using photon counting preceded by small displacement \cite{Banaszek98}. The price to pay is a significant reduction of the repetition rate as it is now given by the rate at which single photons are heralded. A detailed comparison will be presented elsewhere \cite{Caprara14}.\\

As our results give the strategy that optimizes the observed violation, we expect that they will have a significant impact in on-going experiments. Focusing on the experiment reported in Ref. \cite{Christensen13} for example, in which photon pairs distributed over $N=25$ modes and detectors with an (overall) efficiency of $75\%$ have been used for a Bell test while closing the detection loophole, we find the S value $2.0018$ using the optimization of the state structure and measurement settings presented here while the observed value was $2.0002.$ This translates into a speed up in randomness expansion by one order of magnitude. Even if the dark count probability and the fluorescence background are taken into account, we envision a speed up by a factor of $\sim 3$ in the most conservative case. Note that in practice, one is tempted to use a single detector locally, as in Ref. \cite{Christensen13}. In this case, the strategy with which the non-conclusive results are treated is different from the one presented here but we have found that it is also one that is optimal.\\

\paragraph{Acknowledgements.} We warmly thank V. Scarani for helpful discussions and comments. This work was supported by the Swiss NCCR QSIT, the Swiss National Science Foundation SNSF (grant PP00P2$\_$150579 and "Early PostDoc.Mobility"), the European Commission (IP SIQS, Chist-era DIQIP), the Singapore Ministry of Education (partly through the Academic Research Fund Tier 3 MOE2012-T3-1-009) and the Singapore National Research Foundation.\\

\paragraph{Appendix}
We here present a list of all the no-detection probabilities. They provide all the information that is necessary to compute the optimal CHSH value. As far as the notation is concerned, we have maintained the one presented in the main text for $T_g$, $C_g$ ($T_{\bar{g}}$ and $C_{\bar{g}}$), and $C_{\alpha}$ ($C_{\beta}$), while $S_{\alpha}$ ($S_{\beta}$) now means $\sin \alpha$ ($\sin \beta$). Furthermore, $C_{\phi_A-\phi_B}$ means $\cos(\phi_A-\phi_B).$\\
The probability to detect no photon in mode $A$ is given by
\begin{equation}
p(\text{nc}_{A})=(1-p_{\text{dc}})\left(\frac{2}{2-\eta +\eta( C^2_{\alpha} C_{2 g}+ C_{2 \bar{g}} S^2_{\alpha})}\right)^N.
\label{nca}\end{equation}
The expressions for $p(\text{nc}_{A_{\perp}})$, $p(\text{nc}_{B_{\perp}})$, and $p(\text{nc}_{B})$ can be obtained from equation (\ref{nca}) by inverting $g$ and $\bar{g}$, replacing $\alpha$ with $\beta$, and inverting $g$ and $\bar{g}$ and replacing $\alpha$ with $\beta$ respectively.
The probability for no detection in modes $A$ and $A_{\perp}$ ($B$ and $B_{\perp}$) is given by the following expression
\begin{equation}\begin{split}
p&(\text{nc}_{A}\,\&\,\text{nc}_{A_{\perp}})\\
=&p(\text{nc}_{B}\,\&\,\text{nc}_{B_{\perp}})\\
=&(1-p_{\text{dc}})^2\left(\frac{4}{(2-\eta+\eta C_{2 g}) (2-\eta+\eta C_{2\bar{g}})}\right)^N.
\end{split}\end{equation}
The probability for no detection in modes $A$ and $B$ is given by
\begin{equation}\begin{split}
p(\text{nc}_{A}\,\&\,\text{nc}_{B})&=(1-p_{\text{dc}})^24^N(C^2_g C^2_{\bar{g}})^{-N}\times\\
&\Big(4+2\eta ^2 T_g T_{\bar{g}} C_{\phi_A-\phi_B} S_{2 \alpha} S_{2\beta} \\
&-T^2_{\bar{g}}(2-\eta+\eta  C_{2\alpha})(2-\eta-\eta  C_{2 \beta})\\
&+T^2_g ((2-\eta-\eta  C_{2\alpha}) (\eta-2-\eta  C_{2\beta})\\
&+4(1-\eta)^2T^2_{\bar{g}})\Big)^{-N}.
\end{split}\label{nca&ncb}\end{equation}
Similarly, $p(\text{nc}_{A_{\perp}}\,\&\,\text{nc}_{B})$, $p(\text{nc}_{A}\,\&\,\text{nc}_{B_{\perp}})$, and $p(\text{nc}_{A_{\perp}}\,\&\,\text{nc}_{B_{\perp}})$ can be derived from equation (\ref{nca&ncb}) substituting $\alpha$ with $\alpha+\frac{\pi}{2}$, $\beta$ with $\beta+\frac{\pi}{2}$, and $\alpha$ with $\alpha+\frac{\pi}{2}$ and $\beta$ with $\beta+\frac{\pi}{2}$ respectively.
The probability $p(\text{nc}_{A}\,\&\,\text{nc}_{A_{\perp}}\,\&\,\text{nc}_{B})$ of no detection in modes $A$, $A_{\perp}$, and $B$ is given by
\begin{equation}\begin{split}
p&(\text{nc}_{A}\,\&\,\text{nc}_{A_{\perp}}\,\&\,\text{nc}_{B})=(1-p_{\text{dc}})^3\times\\
&\Big(C^2_g C^2_{\bar{g}}(1-\frac{1}{2}(1-\eta) T^2_{ \bar{g}}(2-\eta-\eta C_{2\beta})\\
&-(1-\eta)T^2_{ g}(1-\eta S^2_{\beta}-(1-\eta)^2 T^2_{ \bar{g}}))\Big)^{-N}.
\end{split}\label{nca&ncaperp&ncb}\end{equation}
Similarly the expressions for $p(\text{nc}_{A}\,\&\,\text{nc}_{A_{\perp}}\,\&\,\text{nc}_{B_{\perp}})$, $p(\text{nc}_{A_{\perp}}\,\&\,\text{nc}_{B}\,\&\,\text{nc}_{B_{\perp}})$, and $p(\text{nc}_{A}\,\&\,\text{nc}_{B}\,\&\,\text{nc}_{B_{\perp}})$ can be obtained from equation (\ref{nca&ncaperp&ncb}) inverting $g$ and $\bar{g}$, inverting $\alpha$ and $\beta$, and inverting $g$ and $\bar{g}$ and $\alpha$ and $\beta$ respectively. Finally,
\begin{equation}\begin{split}
p(\text{nc}_{A}&\,\&\,\text{nc}_{A_{\perp}}\,\&\,\text{nc}_{B}\,\&\,\text{nc}_{B_{\perp}})\\
&=(1-p_{\text{dc}})^44^N\times\\
&\Big((1+(1-\eta)^2+(1-(1-\eta)^2)C_{2 g}) \times\\
&(1 +(1-\eta)^2+(1-(1-\eta)^2)C_{2 \bar{g}})\Big)^{-N}.
\end{split}\end{equation}


\begin{thebibliography}{3}

\bibitem{Clauser69} J.F. Clauser, M. Horne, A. Shimony and R.A. Holt, Phys. Rev. Lett. {\bf 23}, 880 (1969).

\bibitem{Clauser74} J. Clauser and M. Horne, Phys. Rev. D {\bf 10}, 526535 (1974).

\bibitem{Brunner13} N. Brunner, D. Cavalcanti, S. Pironio, V. Scarani, S. Wehner, arXiv:1303.2849.

\bibitem{Eberhard93} P.H. Eberhard, Phys. Rev. A {\bf 47}, R747 (1993).

\bibitem{Pironio10} S. Pironio \textit{et al.} Nature {\bf 464}, 1021 (2010).

\bibitem{Christensen13}
B.G. Christensen \textit{et al.} Phys. Rev. Lett. {\bf 111}, 130406 (2013).

\bibitem{Acin07} A. Acin~\textit{et al.}, Phys. Rev. Lett. {\bf 98}, 230501(2007).

\bibitem{Gisin10} N. Gisin, S. Pironio, and N. Sangouard, Phys. Rev. Lett. {\bf 105}, 070501 (2010).

\bibitem{Sangouard11} N. Sangouard~\textit{et al.}, Phys. Rev. Lett. {\bf 106}, 120403 (2011).

\bibitem{Lim13} C.C.W. Lim, C. Portmann, M. Tomamichel, R. Renner, and N. Gisin, Phys. Rev. X {\bf 3}, 031006 (2013).

\bibitem{Rowe01} M.A. Rowe~{\textit et al.}, Nature {\bf 409}, 791 (2001).

\bibitem{Matsukevich08} D.N. Matsukevich {\textit et al.},  Phys. Rev. Lett. {\bf 100}, 1540404 (2008).

\bibitem{Hofmann12} J. Hofmann \textit{et al.}, Science {\bf 337}, 72 (2012).

\bibitem{Aspect82} A. Aspect, J. Dalibard, and G. Roger, Phys. Rev. Lett. {\bf 49}, 1804 (1982).

\bibitem{Tittel98} W. Tittel, J. Brendel, H. Zbinden, and N. Gisin, Phys. Rev. Lett. {\bf 81}, 3563 (1998).

\bibitem{Weihs98} G. Weihs, T. Jennewein, C. Simon, H. Weinfurter, and A. Zeilinger, Phys. Rev. Lett. {\bf 81}, 5039 (1998).

\bibitem{Giustina13}
M. Giustina \textit{et al.} Nature {\bf 497}, 227 (2013).

\bibitem{footnotechi} We assumed that $\chi$ and $\bar \chi$ are independent of $k$, i.e. the distribution of spatial/frequency modes is flat.

\bibitem{Sekatski10}
P. Sekatski, B. Sanguinetti, E. Pomarico, N. Gisin, and C. Simon, Phys. Rev. A {\bf 82}, 053814 (2010).

\bibitem{Collet88} M.J. Collet, Phys. Rev. A {\bf 38}, 2233 (1988).

\bibitem{BohrBrask13} J. Bohr Brask, R. Chaves, and N. Brunner, Phys. Rev. A {\bf 88}, 012111 (2013).

\bibitem{Banaszek98} K. Banaszek and K. Wodkiewicz, Phys. Rev. Lett. {\bf 82}, 2009 (1999).

\bibitem{Caprara14} V. Caprara Vivoli  \textit{et al.}, in preparation.

\bibitem{Massar02} S. Massar and S. Pironio, Phys. Rev. A {\bf 68}, 062109 (2003).

\end{thebibliography}
\end{document}